\documentclass[aps,prf,preprint,superscriptaddress]{revtex4-1}
\usepackage[utf8]{inputenc}
\usepackage{bm}
\usepackage{amssymb,amsmath}
\usepackage{natbib}
\usepackage{graphicx}
\usepackage{diagbox}
\usepackage{siunitx}
\usepackage{soul}       
\usepackage{ulem}       
\usepackage{xcolor}     
\usepackage[running]{lineno} 
\usepackage[caption=false,position=top]{subfig}

\date{May 2020}

\begin{document}

\title{Doing more with less: the flagellar end piece enhances the propulsive effectiveness of human spermatozoa}
\author{Cara V.\ Neal}
\affiliation{School of Mathematics, University of Birmingham, UK}
\affiliation{Centre for Human Reproductive Science, University of Birmingham, UK}
\author{Atticus L.\ Hall-McNair}
\affiliation{School of Mathematics, University of Birmingham, UK}
\affiliation{Centre for Human Reproductive Science, University of Birmingham, UK}
\author{Jackson Kirkman-Brown}
\affiliation{Centre for Human Reproductive Science, University of Birmingham, UK}
\affiliation{Institute for Metabolism and Systems Research, University of Birmingham, UK}
\author{David J.\ Smith}
\email{d.j.smith@bham.ac.uk}
\affiliation{School of Mathematics, University of Birmingham, UK}
\affiliation{Centre for Human Reproductive Science, University of Birmingham, UK}
\affiliation{Institute for Metabolism and Systems Research, University of Birmingham, UK}
\author{Meurig T.\ Gallagher}
\affiliation{School of Mathematics, University of Birmingham, UK}
\affiliation{Centre for Human Reproductive Science, University of Birmingham, UK}
\affiliation{Centre for Systems Modelling and Quantitative Biomedicine, University of Birmingham, UK}
\affiliation{Institute for Metabolism and Systems Research, University of Birmingham, UK}

\begin{abstract}
    Spermatozoa self-propel by propagating bending waves along a predominantly active elastic flagellum. 
    The organized structure of the ``9 + 2'' axoneme is lost in the most-distal few microns of the flagellum, and therefore this region is unlikely to have the ability to generate active bending; as such it has been largely neglected in biophysical studies. 
    Through elastohydrodynamic modeling of human-like sperm we show that an inactive distal region confers significant advantages, both in propulsive thrust and swimming efficiency, when compared with a fully active flagellum of the same total length. 
    The beneficial effect of the inactive end piece on these statistics can be as small as a few percent but can be above 430\%. 
    The optimal inactive length, between 2--18\% of the total length, depends on both wavenumber and viscous-elastic ratio, and therefore is likely to vary in different species. Potential implications in evolutionary biology and clinical assessment are discussed.
\end{abstract}
\maketitle
\newpage
\section{Introduction}
Spermatozoa, alongside their crucial role in sexual reproduction, are a principal motivating example of inertialess propulsion in the very low Reynolds number regime. The time-irreversible motion required for effective motility is achieved through the propagation of bending waves along the eukaryotic axoneme, which consists of 9 doublet microtubules (dMTs) surrounding a central pair (CP) (see Fig.~\ref{fig:sperm-schematic}a). This ``9+2'' axoneme forms the active elastic internal core of the slender flagellum, and in the case of human sperm is surrounded by a fibrous sheath which thins towards the distal end. While sperm morphology varies significantly between species \cite{austin1995evolution,fawcett1975mammalian,werner2008insect,mafunda2017sperm,nelson2010tardigrada, anderson1975form}, there are clear conserved features which can be seen in humans, most mammals, and also our evolutionary ancestors \cite{cummins1985mammalian}. In gross structural terms, sperm comprise 
(i) the head, which contains the genetic cargo;
(ii) the midpiece of the flagellum, typically a few microns in length, containing the ATP-generating mitochondria;
(iii) the principal piece of the flagellum, typically 40--60\(\,\mu\)m in length (although much longer in some species \cite{bjork2006intensity}), the core of which is the 9+2 axoneme that produces and propagates active bending waves through dynein-ATPase activity \cite{machin1958wave}; and (iv) the end piece, typically a few microns in length, which consists of disorganized singlet microtubules (sMTs) only \cite{zabeo2019axonemal}. Lacking the predominant 9+2 axonemal structure it appears unlikely that the end piece is a site of molecular motor activity, a hypothesis that forms the basis for our study. Since the end piece is assumed to be unactuated, we will refer to it as \textit{inactive}, noting however that this does not mean it is necessarily \textit{ineffective}. Correspondingly, the actuated principal piece will be referred to as \textit{active}.
Detailed review of human sperm morphology can be found in \cite{gaffney2011mammalian,lauga2009hydrodynamics}.

While the end piece can be observed through transmission electron and atomic force microscopy \cite{fawcett1975mammalian,ierardi2008afm,lin2018asymmetric}, live imaging to determine its role in cell propulsion is currently challenging. 
Furthermore, because the end piece has been largely considered to not have a role in propelling the cell, it has received relatively little attention. 
However, analysis of experimental data \cite{gallagher2019rapid} indicates that the sperm flagellar waveform has a significant impact on propulsive effectiveness, and moreover changes to the waveform have an important role in enabling cells to penetrate the highly viscous cervical mucus encountered in internal fertilization \cite{smith2009bend}.
This leads us to ask: \textit{does the presence of a mechanically inactive region at the end of the flagellum help or hinder the cell's progressive motion?}
\section{Model setup}

\begin{figure}[t]
    \centering
    \includegraphics[width=\textwidth,trim={0cm 0cm 0cm 0cm}]{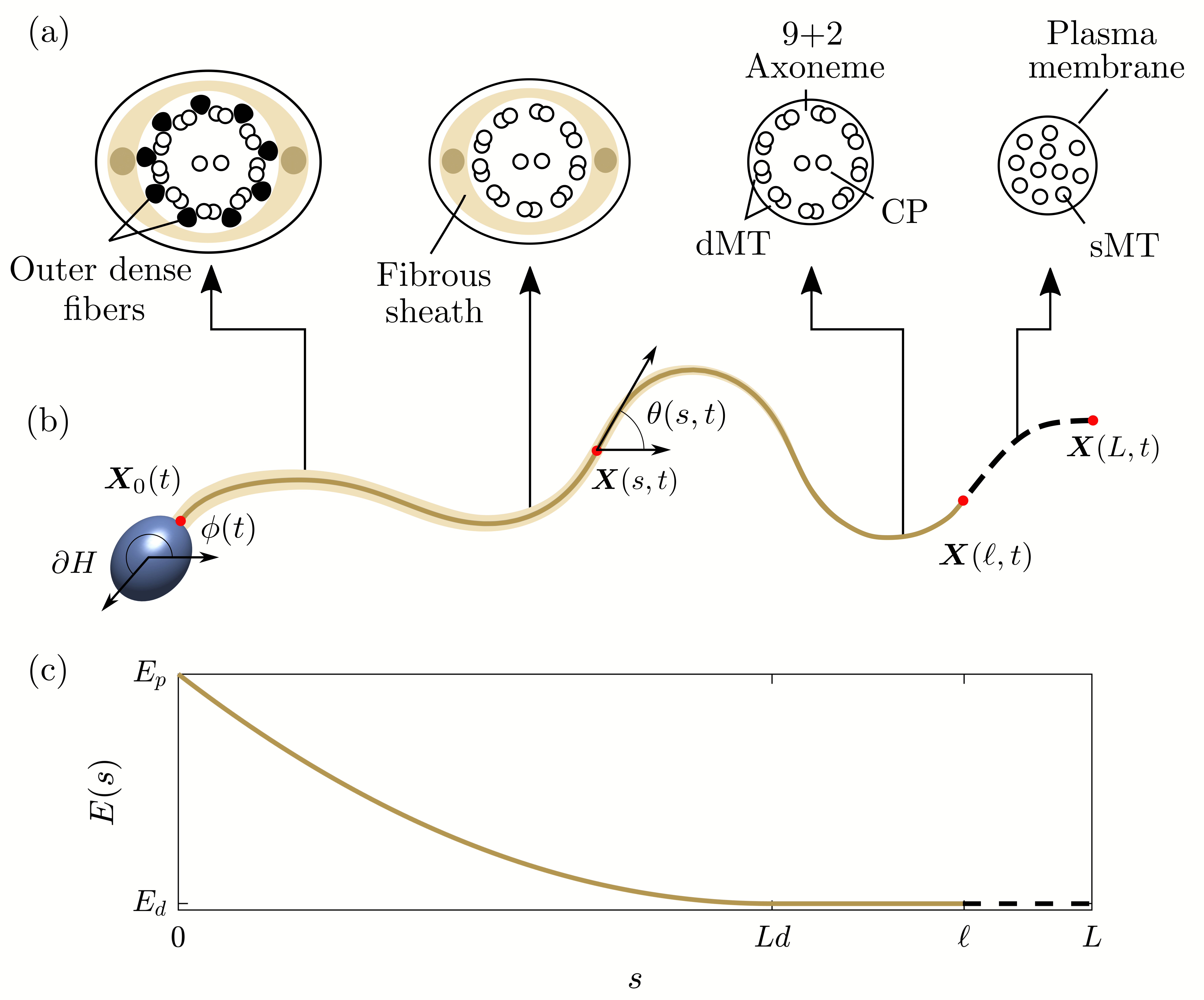}
    \caption{Schematic of an idealized sperm cell. (a) Cross sections (redrawn from \cite{zabeo2019axonemal}) showing internal flagellar structure, where sMT denotes singlet microtubules, dMT doublet microtubules and CP the central microtubule pair. 
    (b) A sketch of the model sperm highlighting the position of the centerline \(\bm{X}(s,t)\) (parameterized by arclength \(s\in[0,L]\) and tangent angle \(\theta(s,t)\)), head surface (\(\partial H\), blue ellipsoid), head angle (\(\phi(t)\)), and flagellum/neck junction (\(\bm{X}_0(t)\)). The active component of the flagellum (\(s \in [0,\ell]\)) is shown in dark yellow, with the inactive distal end (\(s \in [\ell,L]\)) shown black dashed. 
    (c) A tapering fibrous sheath surrounds the core of the flagellum, modeled through the varying elastic stiffness function (Eq.~\eqref{eq:stiffness}).}
    \label{fig:sperm-schematic}
\end{figure}

The emergence of elastic waves on the flagellum can be described by a constitutively linear, geometrically nonlinear filament, with the addition of an active moment per unit length \(m\), which models the internal sliding produced by dynein activity, and a hydrodynamic term \(\bm{f}\) which describes the force per unit length exerted by the filament onto the fluid. 
Many sperm have approximately planar waveforms, especially upon approaching and collecting at surfaces \cite{gallagher2018casa,woolley2003motility}. 
As such, their shape can be fully described by the angles made between the tangent and the lab frame horizontal (denoted  \(\theta(s,t)\)), and between the head centerline angle and the lab frame horizontal (denoted \(\phi(t)\)), as shown in Fig.~\ref{fig:sperm-schematic}b. 
The sperm head is modeled by an ellipsoidal surface $\partial H$, with the head-flagellum joint denoted by $\bm{X}_0(t)$.
Following \cite{moreau2018asymptotic,hall2019efficient}, we parameterize the filament by arclength \(s\), with \(s=0\) corresponding to the head-flagellum joint and \(s=L\) to the distal end of the flagellum, and apply force and moment free boundary conditions at \(s=L\) to get
\begin{equation}
    E(s)\,\partial_s \theta(s,t) - \bm{e}_3\cdot\int_s^{L} \partial_{s'} \bm{X}(s',t) \times \left(\int_{s'}^{L} \bm{f}(s'',t)\, ds''\right) ds' - \int_s^{L} m(s',t)\,ds' =0, \label{eq:elasticity0}
\end{equation}
with the elastic stiffness given, using a modification of previous work (\cite{gaffney2011mammalian}), by
\begin{equation}
E(s)=
\begin{cases}
E_d(\rho-1)\left( \frac{s-Ld}{Ld}\right)^2 + E_d & s \leq Ld, \\
E_d  & s>Ld,
\end{cases}
\label{eq:stiffness}
\end{equation}
where \(E_d\) is the stiffness of the distal flagellum. The dimensionless parameters \(\rho\), corresponding to proximal-distal stiffness ratio, and \(d\), corresponding to the proportion of the flagellum over which the stiffness is a varying function, will be fixed throughout this study; \(d = 0.65\) is chosen so that \(Ld=39\,\)\(\mu\)m for human sperm, corresponding to the distance over which the accessory structures extend, while \(\rho \approx 35\) is approximated based on a qualitative comparison with experimental data (Appendix~\ref{app:stiff}). A sketch of how the elastic stiffness $E(s)$ varies with arclength $s$ can be seen in Fig.~\ref{fig:sperm-schematic}c. The effect of varying the remaining parameters of wavelength, active length, total length, viscosity, distal stiffness and frequency is investigated via analyzing a dimensionless model. 

Returning to Eq.~(\ref{eq:elasticity0}), the position vector \(\bm{X}=\bm{X}(s,t)\) describes the flagellar waveform at time \(t\), so that \(\partial_s\bm{X}\) is the tangent vector, and \(\bm{e}_3\) is a unit vector pointing perpendicular to the plane of beating. Integrating by parts leads to the elasticity integral equation
\begin{equation}
    E(s)\,\partial_s \theta(s,t) + \bm{e}_3\cdot\int_s^{L} (\bm{X}(s',t)-\bm{X}(s,t)) \times \bm{f}(s',t) \, ds' - \int_s^{L} m(s',t)\,ds' =0. \label{eq:elasticity}
\end{equation}

The active moment density can be described to a first approximation by a sinusoidal traveling wave \(m(s,t)=m_0 \cos(k s-\omega t)\), where \(k\) is wavenumber and \(\omega\) is radian frequency. The inactive end piece can be modeled by taking the product with a Heaviside function, so that \(m(s,t) = m_0 \cos(k s-\omega t)H(\ell -s)\) where \(0<\ell\leqslant L\) is the length of the active tail segment.

At very low Reynolds number, neglecting non-Newtonian influences on the fluid, the hydrodynamics are described by the Stokes flow equations
\begin{equation}
    -\bm{\nabla}p + \mu \nabla^2\bm{u} = \bm{0}, \quad \bm{\nabla}\cdot \bm{u} = 0,
\end{equation}
\sloppy
where \(p=p(\bm{x},t)\) is pressure, \(\bm{u}=\bm{u}(\bm{x},t)\) is velocity and \(\mu\) is dynamic viscosity. These equations are augmented by the no-slip, no-penetration boundary condition \({\bm{u}(\bm{X}(s,t),t)=\partial_t\bm{X}(s,t)}\), i.e.\ the fluid in contact with the filament moves at the same velocity as the filament. A convenient and accurate numerical method to solve these equations for biological flow problems with deforming boundaries is based on the `regularized stokeslet' \cite{cortez2001method,cortez2005method}, i.e.\ the solution to the exactly incompressible Stokes flow equations driven by a spatially-concentrated but smoothed force
\begin{equation}
    -\bm{\nabla}p + \mu \nabla^2\bm{u} + \psi_\varepsilon(\bm{x},\bm{y})\bm{e}_3 = 0, \quad \bm{\nabla}\cdot \bm{u} = 0,
\end{equation}
where \(\varepsilon\ll 1\) is a regularization parameter, $\bm{y}$ is the location of the force, $\bm{x}$ is the evaluation point and \(\psi_\varepsilon\) is a smoothed approximation to a Dirac delta function. The choice 
\begin{equation}
    \psi_\varepsilon(\bm{x},\bm{y})=15\varepsilon^4/r_\varepsilon^{7},
\end{equation}
leads to the regularized stokeslet \cite{cortez2005method}
\begin{equation}
      S_{ij}^\varepsilon(\bm{x},\bm{y})=\frac{1}{8\pi\mu}\left(\frac{\delta_{ij}(r^2+2\varepsilon^2)+r_ir_j}{r_\varepsilon^3} \right) ,  
    \end{equation}
where \(r_i=x_i-y_i\), \(r^2=r_i r_i\), \(r_\varepsilon^2=r^2+\varepsilon^2\).

The flow \(u_j(\bm{x},t)\) produced by a filament \(\bm{X}(s,t)\) exerting force per unit length \(\bm{f}(s,t)\) is then given by the line integral \(\int_0^{L} S_{jk}^\varepsilon(\bm{x},\bm{X}(s,t))f_k(s,t)\,ds\). The flow due to the surface of the sperm head \(\partial H\), exerting force per unit area \(\bm{\varphi}(\bm{Y},t)\) for \(\bm{Y}\in\partial H\), is given by the surface integral \(\iint_{\partial H} S_{jk}^\varepsilon(\bm{x},\bm{Y})\varphi_k(\bm{Y})\,dS_{\bm{Y}}\), yielding the boundary integral equation \cite{smith2009boundaryelement} for the hydrodynamics, namely 
\begin{equation}
    u_j(\bm{x},t)=\int_0^{L} S^\varepsilon_{jk}(\bm{x},\bm{X}(s,t))f_k(s,t)\,ds+\iint_{\partial H} S_{jk}^\varepsilon(\bm{x},\bm{Y})\varphi_k(\bm{Y},t)\,dS_{\bm{Y}}. \label{eq:flow}
\end{equation}

The position and shape of the cell can be described by the location \(\bm{X}_0(t)\) of the head-flagellum joint and the waveform \(\theta(s,t)\), so that the flagellar curve is 
\begin{equation}
    \bm{X}(s,t)=\bm{X}_0(t)+\int_0^s [\cos\theta(s',t),\sin\theta(s',t),0]^T \,ds'. \label{eq:geometry}
\end{equation}
Differentiating with respect to time, the flagellar velocity is then given by
\begin{equation}
    \bm{u}(\bm{X}(s,t),t)=\dot{\bm{X}}_0(t)+\int_0^s\partial_t\theta(s',t)[-\sin\theta(s',t),\cos\theta(s',t),0]^T \,ds'. \label{eq:kinematic1}
\end{equation}
Modeling the head as undergoing rigid body motion about the head-flagellum joint, the surface velocity of a point \(\bm{Y}\in\partial H\) is given by
\begin{equation}
    \bm{u}(\bm{Y}(t),t)=\dot{\bm{X}}_0(t)+\partial_t \phi(t)\,\bm{e}_3 \times (\bm{Y}(t)-\bm{X}_0(t)). \label{eq:kinematic2}
\end{equation}
Equations.~\eqref{eq:kinematic1} and \eqref{eq:kinematic2} couple with fluid mechanics (Eq.~\eqref{eq:flow}), active elasticity (Eq.~\eqref{eq:elasticity}), and total force and moment balance across the cell to yield a model for the unknowns \(\theta(s,t)\), \(\phi(t)\), \(\bm{X}_0(t)\), \(\bm{f}(s,t)\) and \(\bm{\varphi}(\bm{Y},t)\). Nondimensionalizing with lengthscale \(L\), timescale \(1/\omega\) and force scale \(\mu\omega L^2\) yields the elasticity integral equation in scaled variables (with dimensionless variables denoted by $\,\hat{}\,$ )
\begin{linenomath}
\begin{multline}
    E(\hat{s})\,\partial_{\hat{s}} \theta(\hat{s},\hat{t})+ \bm{e}_3\cdot\mathcal{S}^4\int_{\hat{s}}^1 (\hat{\bm{X}}(\hat{s}',\hat{t})\,-\hat{\bm{X}}(\hat{s},\hat{t})) \times \hat{\bm{f}}(\hat{s}',\hat{t})  \,d\hat{s}' \\- \mathcal{M}\mathcal{S}^4\int_{\hat{s}}^1 \cos(\hat{k}\hat{s}'-\hat{t})H(\hat{\ell}-\hat{s}') \,d\hat{s}' =0, \label{eq:elasticityND}
\end{multline}
\end{linenomath}
where \(\mathcal{S}=L(\mu\omega/E_d)^{1/4}\) is a dimensionless group comparing viscous and elastic forces (related, but not identical to, the commonly-used `sperm number'), \(\mathcal{M}=m_0/\mu\omega L^2\) is a dimensionless group comparing active and viscous forces, and $\hat{\ell}=\ell/L$ is the dimensionless length of the active segment. Here, $E_d$ is the stiffness at the distal tip of the flagellum ($\hat{s}=1$) and the dimensionless wavenumber is \(\hat{k}=k L\). The remaining equations nondimensionalize directly using these scales.

The problem is numerically discretized as described by Hall-McNair \textit{et al}. \cite{hall2019efficient}, accounting for nonlocal hydrodynamics via the method of regularized stokeslets \cite{cortez2005method}. This framework is
modified to take into account the presence of the head via the nearest-neighbor discretization of Gallagher \& Smith \cite{gallagher2018meshfree}. The head-flagellum coupling is enforced via the dimensionless moment balance boundary condition
\begin{equation}
    \hat{\kappa}(0,\hat{t})-\bm{e}_3\cdot\mathcal{S}^4\iint_{\partial \hat{H}}(\hat{\bm{Y}}(\hat{t}\,)-\hat{\bm{X}}_0(\hat{t}\,))\times\hat{\bm{\varphi}}(\hat{\bm{Y}},\hat{t}\,)\,dS_{\hat{\bm{Y}}}=0,
\end{equation}
where the dimensionless curvature evaluated at the head-flagellum joint \(\hat{\kappa}(0,\hat{t})\) is calculated as the centered difference between the head angle \(\phi(\hat{t})\) and the tangent angle of the first segment of the flagellum \(\theta_1(\hat{t})\).

For the remainder of this paper we work with the dimensionless model, but for readability omit the $\,\hat{}\,$ notation used to represent dimensionless variables. The initial value problem for the trajectory, head angle, discretized waveform and force distributions is solved in MATLAB\textsuperscript{\textregistered} using the built-in solver $\mathtt{ode23tb}$. At any point in time, the sperm cell's position and shape can be reconstructed completely from $\bm{X}_0(t)$, $\theta(s,t)$ and \(\phi(t)\) through Eq.~(\ref{eq:geometry}). Simulated sperm cells are initialized with a flagellar shape in the form of a low-amplitude parabola, obtained by sampling a section of unit arclength from the curve \(y = 0.1 x^2\) centered about \(x = 0\).

In what follows, we consider how varying the three dimensionless groups \(\mathcal{S}\), \(\mathcal{M}\), and \(\ell\) through physiological ranges can affect both swimming velocity and efficiency of simulated spermatozoa.
\section{Results}
\begin{figure}[t]
    \centering
    \includegraphics[width=\textwidth]{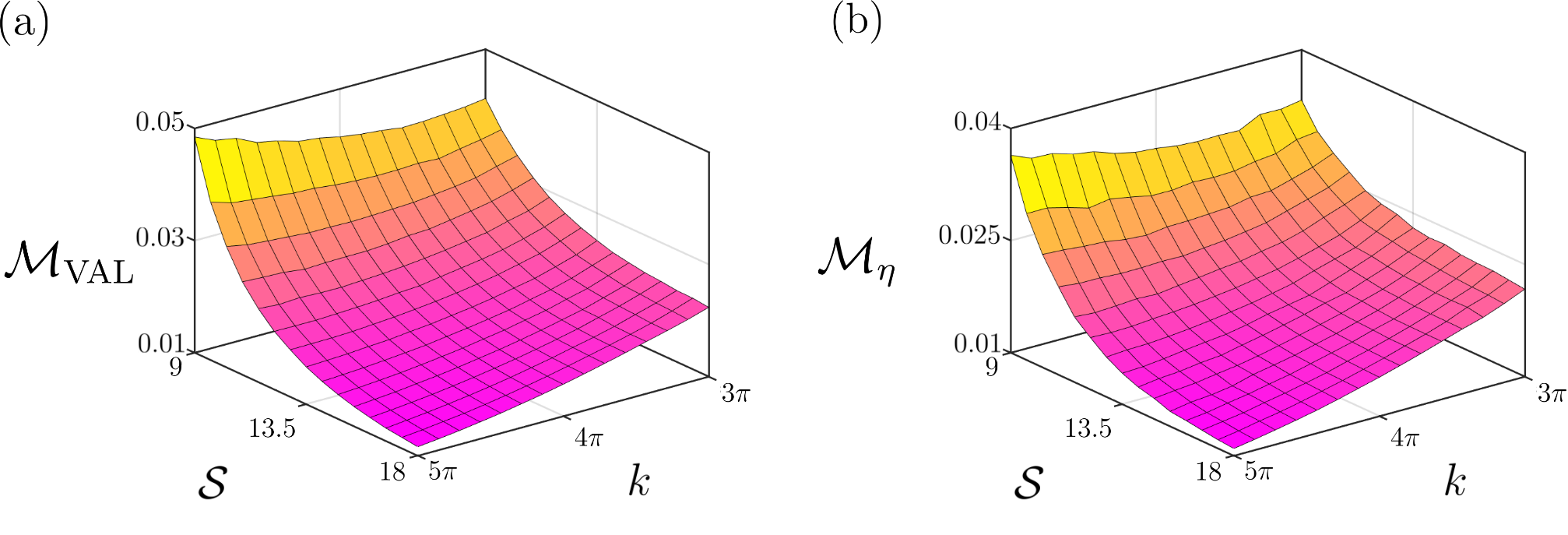}
    \caption{
         Values of the actuation parameter \(\mathcal{M}\), optimizing (a) swimming speed (VAL) and (b) the Lighthill efficiency (\(\eta\)) for \(\mathcal{S}\in[9,18]\) and \(k\in[3\pi,5\pi]\).
    }
    \label{fig:Mval_Meta}
\end{figure}
\begin{figure}[t]
    \centering
    \includegraphics[width=\textwidth]{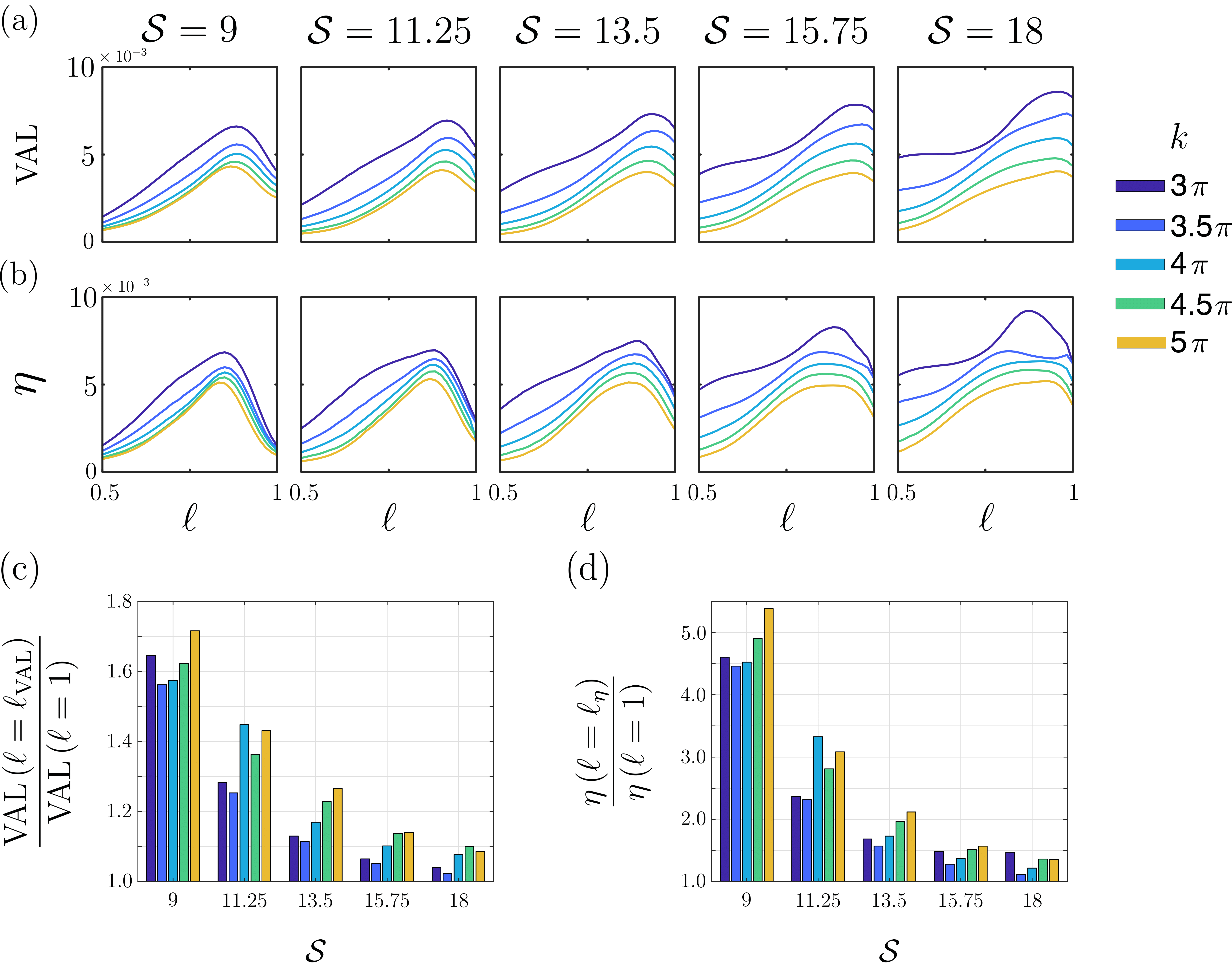}
    \caption{
        The effect of the inactive end piece length on swimming speed and efficiency of propulsion, for simulations actuated with \(\mathcal{M} = \mathcal{M}_{\text{VAL}}\) (a) The swimming speed (VAL), (b) the Lighthill efficiency (\(\eta\)). (c,\,d) The relative increase in VAL and efficiency respectively, when comparing the optimally-active and fully-active flagella, for viscous-elastic parameter choices \(\mathcal{S}\in[9,18]\) and wavenumbers \(k\in [3\pi,5\pi]\).
    }
    \label{fig:results-main}
\end{figure}
The impact of the length of the inactive end piece on propulsion is quantified by the swimming speed and efficiency. Velocity along a line (VAL) is used as a measure of swimming speed, calculated via
\begin{equation}
    \text{VAL}^{(j)} = \|\bm{X}_0^{(j)}-\bm{X}_0^{(j-1)}\| / T ,
\end{equation}
where $T=2\pi$ is the period of the driving wave and $\bm{X}_0^{(j)}$ represents the position of the head-flagellum joint after $j$ periods. Lighthill efficiency \cite{lighthill1975mathematical} is calculated as
\begin{equation}
    \eta^{(j)} = \left(\text{VAL}^{(j)}\right)^2 / \,\overline{W}^{(j)},
\end{equation}
where $\overline{W}^{(j)} = \left< \int_{0}^{1}\bm{u} \cdot \bm{f} ds' + \iint_{\partial H}\bm{u}(\bm{Y})\cdot \bm{\varphi}(\bm{Y})\mathrm{d}S_{\bm{Y}} \right>$ is the average work done by the cell over the $j$\textsuperscript{th} period. In the following, \( j \) is chosen sufficiently large so that the cell has established a regular beat before its statistics are calculated ($j=5$ is sufficient for what follows).

\subsection{Choice of parameters}

\begin{figure}[t]
    \centering
    \includegraphics[width=\textwidth]{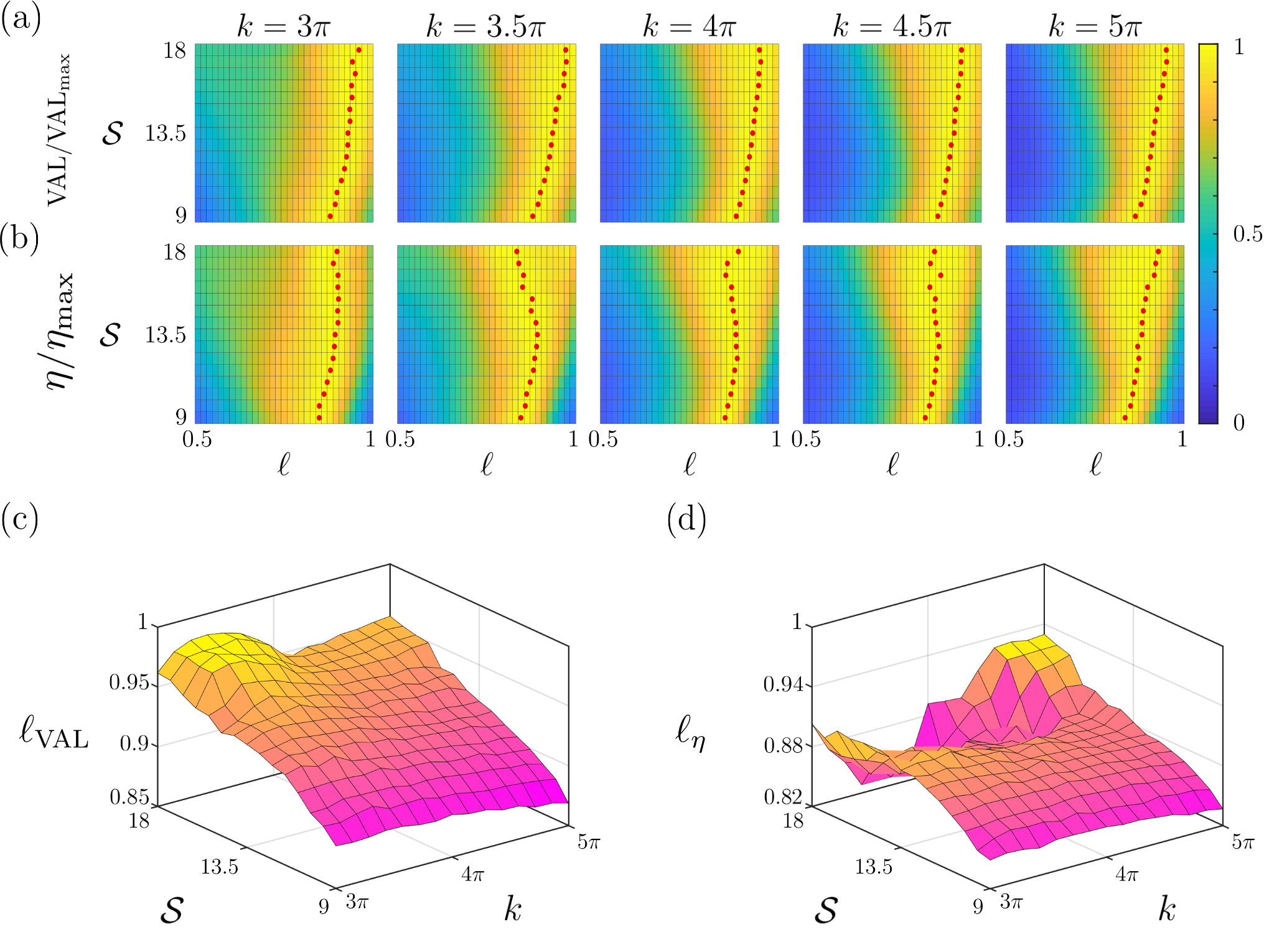}
    \caption{
        (a) Normalized VAL and (b) normalized Lighthill efficiency (\(\eta\)) for \(\mathcal{S}\in [9,18]\) versus \(\ell\in[0.5,1]\), shown for five choices of dimensionless wavenumber \( k \). Values in each subplot are normalized with respect to either (a) the maximum VAL or (b) the maximum \( \eta \) for each \( (\mathcal{S},k)\) pair. The optimal choices of $\ell$ are highlighted as red dots. (c) The active length that optimizes VAL (\(\ell_{\text{VAL}}\)) and (d) the active length that optimizes the Lighthill efficiency ($\ell_\eta$), for \(\mathcal{S}\in [9,18]\), \(k\in [3\pi,5\pi]\). Simulations are actuated with \(\mathcal{M} = \mathcal{M}_{\text{VAL}}\).
    } 
    \label{fig:optimallength}
\end{figure}

\begin{figure}[t]
    \centering
    \includegraphics[width=\textwidth]{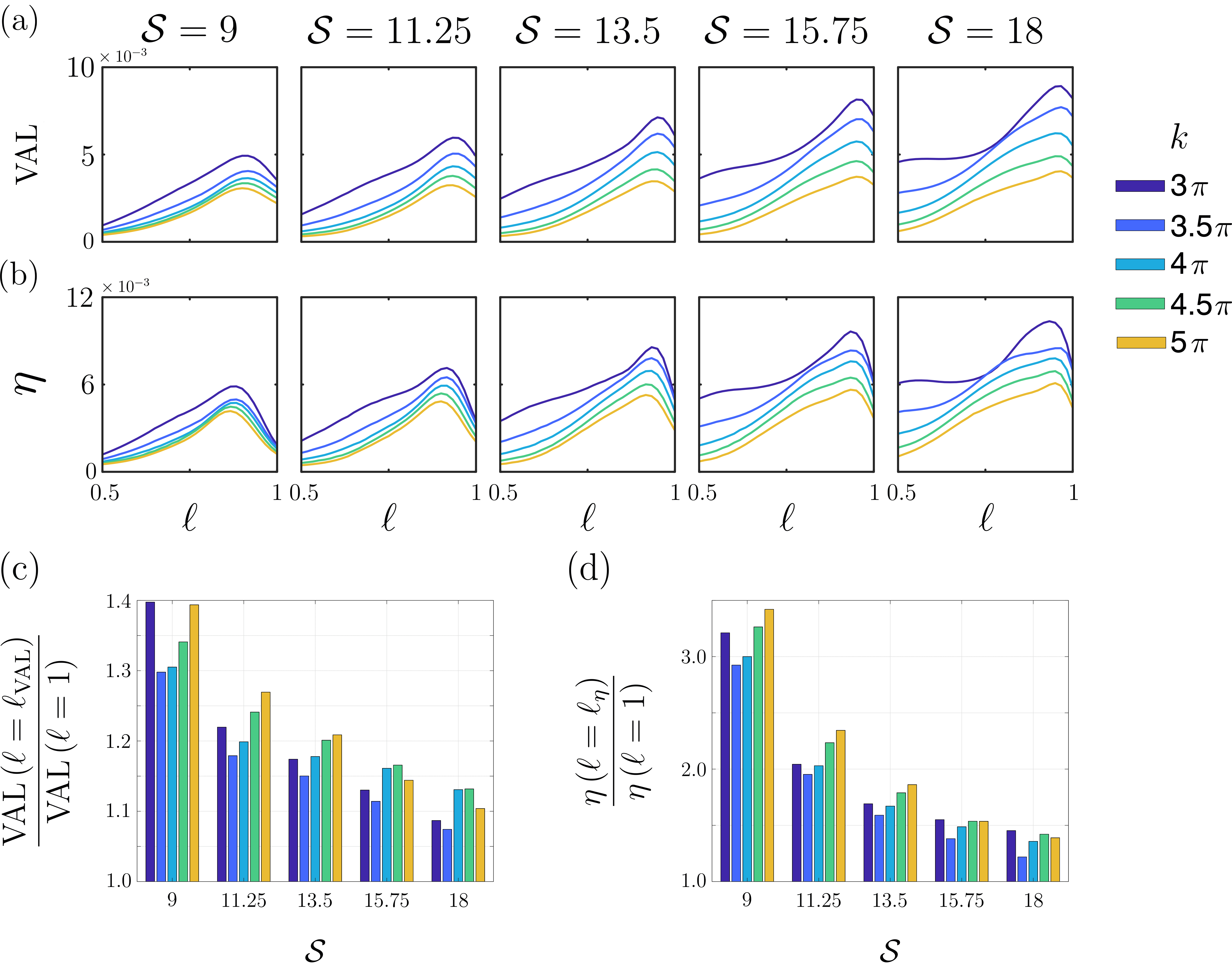}
    \caption{
        The effect of the inactive end piece length on swimming speed and efficiency of propulsion, for simulations actuated with \(\mathcal{M} = \mathcal{M}_{\eta}\) (a) The swimming speed (VAL), (b) the Lighthill efficiency (\(\eta\)). (c,\,d) The relative increase in VAL and efficiency respectively, when comparing the optimally-active and fully-active flagella, for viscous-elastic parameter choices \(\mathcal{S}\in[9,18]\) and wavenumbers \(k\in [3\pi,5\pi]\).
    }
  \label{fig:val_eta_Meta}
  \end{figure}
Below, the viscous-elastic parameter \(\mathcal{S}\) is varied between \(9\) and \(18\), corresponding to the approximate physiological range for human spermatozoa migrating within the female reproductive tract. However, the sperm of many other mammalian species exhibit similar values of \(\mathcal{S}\) including bull (\(\mathcal{S} \approx 10.5\)), chocolate wattled bat (\(\mathcal{S} \approx 12.6\)), rabbit (\(\mathcal{S} \approx 14.85\)), cat (\(\mathcal{S} \approx 16.4\)), dog (\(\mathcal{S} \approx 16.9\)), and dolphin (\(\mathcal{S} \approx 18\)) as well as many others \cite{cummins1985mammalian}.
 
For each \((\mathcal{S},k)\), two variations on the actuation parameter $\mathcal{M}$ are investigated: the value  $\mathcal{M}_\text{VAL}$ that optimizes VAL, and the value \(\mathcal{M}_{\eta}\) that optimizes Lighthill efficiency \(\eta\), each for in the fully-active \(\ell=1\) case. Therefore any increase in velocity or efficiency as a consequence of varying \(\ell\) is \textit{not} a consequence of the specific value of \(\mathcal{M}\) chosen. Indeed any increases in velocity and efficiency observed will therefore be lower bounds for what is possible if \(\mathcal{M}\) is allowed to vary freely. Optimization is carried out via the MATLAB\textsuperscript{\textregistered} 1-dimensional optimization algorithm {\texttt{fminbnd}}. Figure~\ref{fig:Mval_Meta} shows the smooth variation in both \(\mathcal{M}_{\text{VAL}}\) and \(\mathcal{M}_{\eta}\) as \(\mathcal{S}\) and \(k\) are varied.

\subsection{Simulations actuated with $\mathcal{M} =\mathcal{M}_\text{VAL}$} \label{sec:results_mval}
  \begin{figure}[t]
    \centering
    \includegraphics[width=\textwidth]{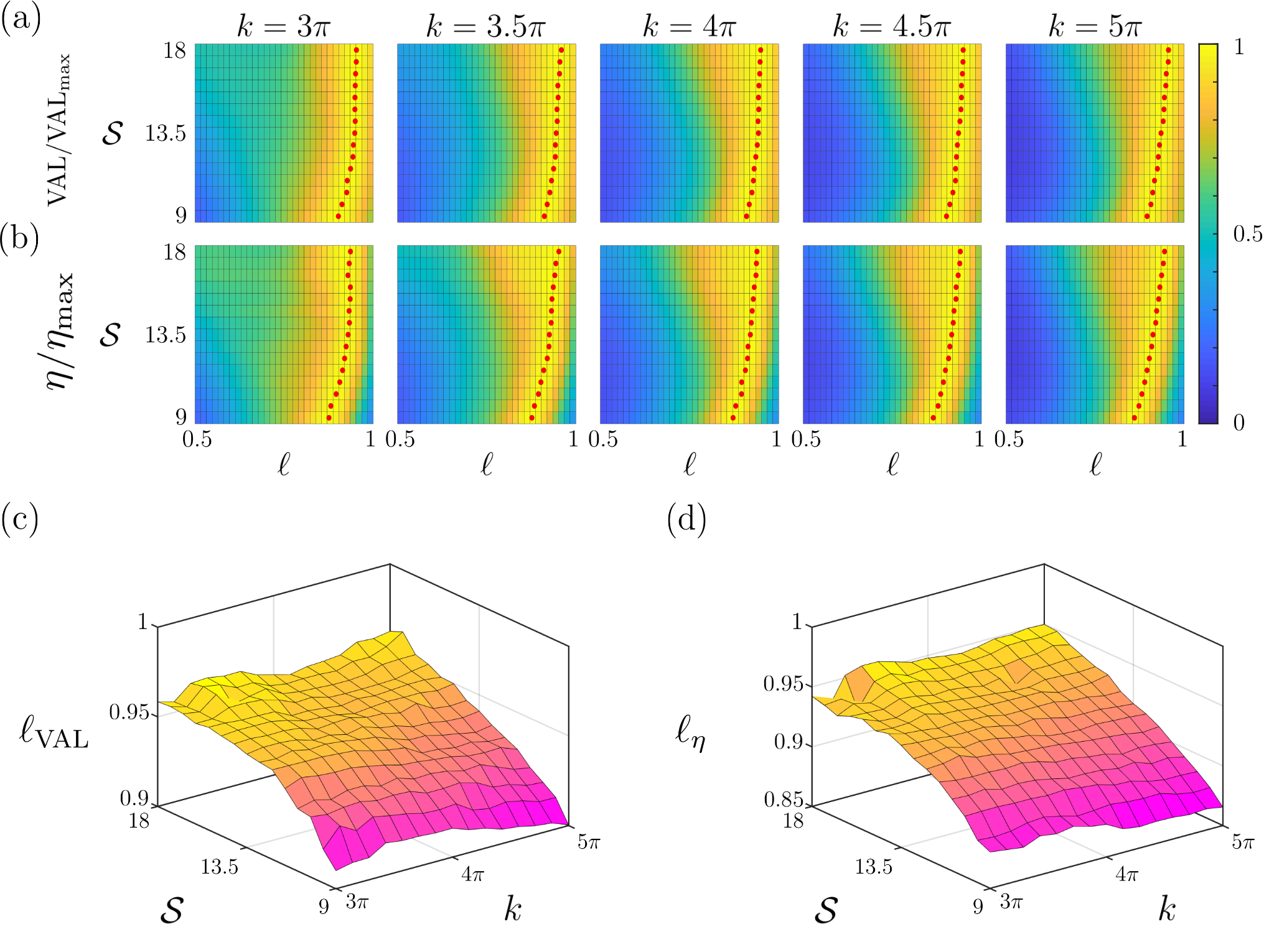}
    \caption{
        (a) Normalized VAL and (b) normalized Lighthill efficiency (\(\eta\)) for \(\mathcal{S}\in [9,18]\) versus \(\ell\in[0.5,1]\), shown for five choices of dimensionless wavenumber \( k \). Values in each subplot are normalized with respect to either (a) the maximum VAL or (b) the maximum \( \eta \) for each \( (\mathcal{S},k)\) pair. The optimal choices of $\ell$ are highlighted as red dots. (c) The active length that optimizes VAL (\(\ell_{\text{VAL}}\)) and (d) the active length that optimizes the Lighthill efficiency ($\ell_\eta$), for \(\mathcal{S}\in [9,18]\), \(k\in [3\pi,5\pi]\). Simulations are actuated with \(\mathcal{M} = \mathcal{M}_\eta\).
    } 
   \label{fig:optimallength_Meta}
\end{figure}
The effects of varying the dimensionless active tail length on sperm swimming speed and efficiency are shown in Figs.~\ref{fig:results-main} for five choices of dimensionless wavenumber \(k\) and viscous-elastic parameter $\mathcal{S}$, actuated with $\mathcal{M}=\mathcal{M}_\text{VAL}$. 
Here \( \ell=1 \) corresponds to an entirely active flagellum and \( \ell = 0 \) to an entirely inactive flagellum. Values \( 0.5 \leqslant \ell \leqslant 1 \) are considered so that the resulting simulations produce cells that are likely to be biologically realistic. Higher wavenumbers are considered as they are typical of mammalian sperm flagella in higher viscosity media \cite{smith2009bend}. 

\begin{figure}[t]
    \centering
    \includegraphics[width=\textwidth]{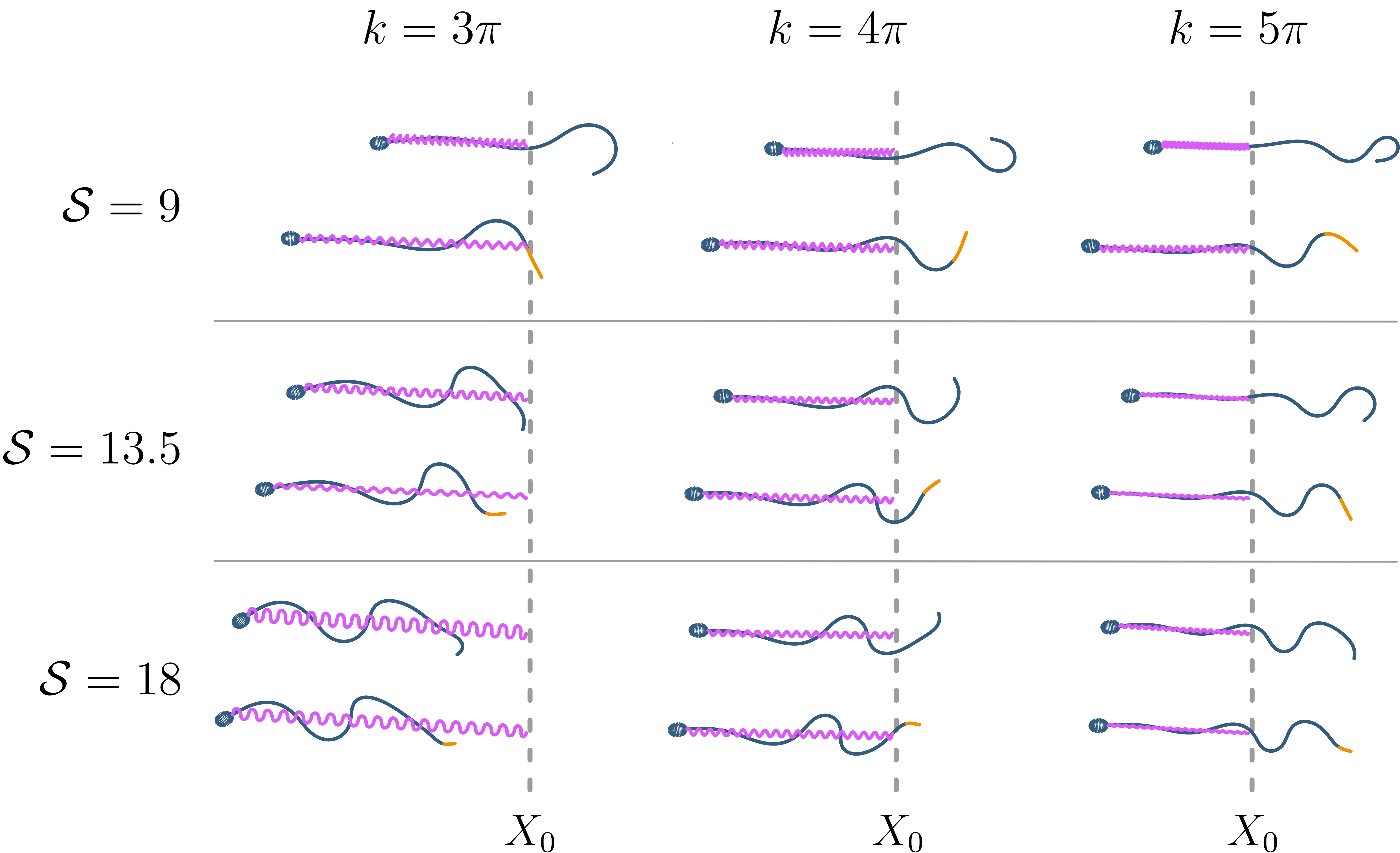}
    \caption{
         The swimming track following the head for both fully-active and optimally-inactive sperm (top and bottom of each pair respectively) for a selection of parameter pairs (\(\mathcal{S},k\)) actuated with \(\mathcal{M}=\mathcal{M}_\text{VAL}\) and simulated for 10 cycles of the active moment. In each pair the sperm is plotted at the final time point with the optimally-inactive length \(\ell=\ell_{\text{VAL}}\), shown in orange.
    }
    \label{fig:tracks}
    \end{figure}
Optimal active lengths for swimming speed, $\ell_{\text{VAL}}$, and efficiency, $\ell_{\eta}$, occur for each parameter pair $(\mathcal{S},k)$; crucially, as shown in Figs.~\ref{fig:results-main}a and \ref{fig:results-main}b, the optima are always less than 1, indicating that by either measure some length of inactive flagellum is always better than a fully active flagellum. 
The relative effect of the end piece on \(\text{VAL}\) and \(\eta\) is larger for smaller values of \(\mathcal{S}\). In particular, for \(\mathcal{S}=9\) the optimally-inactive flagellum results in an 56--72\% increase in \(\text{VAL}\) over \(k\in[3\pi,5\pi]\) compared to a fully active flagellum (Fig.~\ref{fig:results-main}c), decreasing to 2--9\% when $\mathcal{S}=18$. The end piece has a more pronounced effect on the efficiency of cells (Fig.~\ref{fig:results-main}d), with an 11--438\% increase over the values of $k$ and $\mathcal{S}$ considered. 
\begin{figure}[t]
    \centering
    \includegraphics[width=\textwidth]{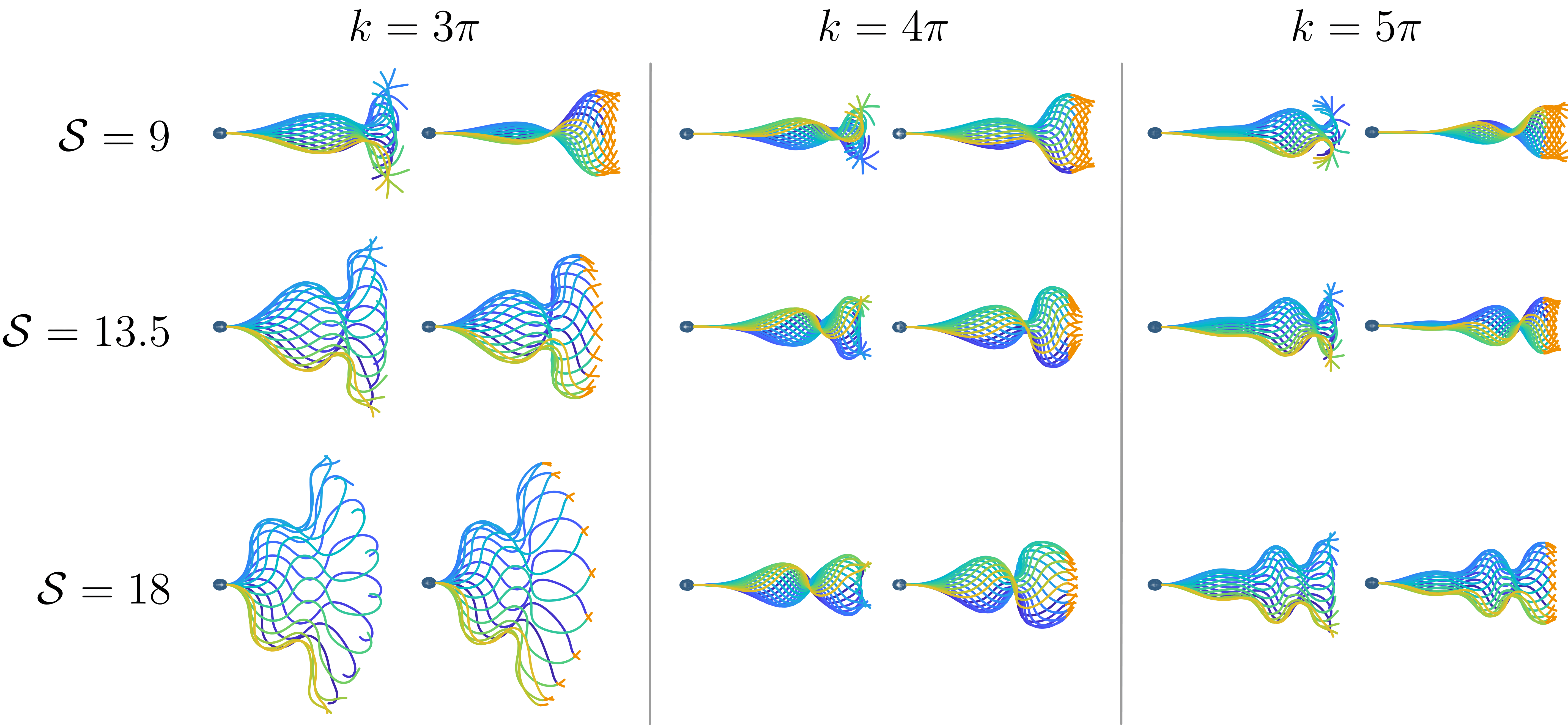}
    \caption{
         The flagellar waveform for a fully-active and optimally-inactive sperm (left and right of each pair respectively), simulated for a selection of parameter pairs (\(\mathcal{S},k\)), and actuated with \(\mathcal{M} = \mathcal{M}_{\text{VAL}}\). The color of the flagellum indicates progression through a single beat, from blue to yellow. For each pair the sperm are plotted with the optimally-inactive length \(\ell_{\text{VAL}}\) shown in orange.
    }
    \label{fig:waveforms}
\end{figure}
A more detailed investigation of the relationship between optimum active flagellum length and each of VAL and \(\eta\) is shown in Figs.~\ref{fig:optimallength}a\,b by simulating cells over a finer gradation in $\mathcal{S}\in[9,18]$. The optimum values \( \ell_{\text{VAL}} \) and \( \ell_\eta \) are shown in Figs.~\ref{fig:optimallength}c and \ref{fig:optimallength}d; typically \( \ell_{\text{VAL}} \neq \ell_{\eta} \) for a given swimmer. For each metric, the optimum active length is smoothly varying for each choice of $\mathcal{S}$ and $k$. In the case of \(\ell_\eta\), we observe non-monotonic behavior for the combination of larger values of \(\mathcal{S}\) and smaller values of \(k\), due to the development of a skewed beat pattern (see Fig.~\ref{fig:tracks}). 

\begin{figure}[t]
    \centering
    \includegraphics[width=\textwidth]{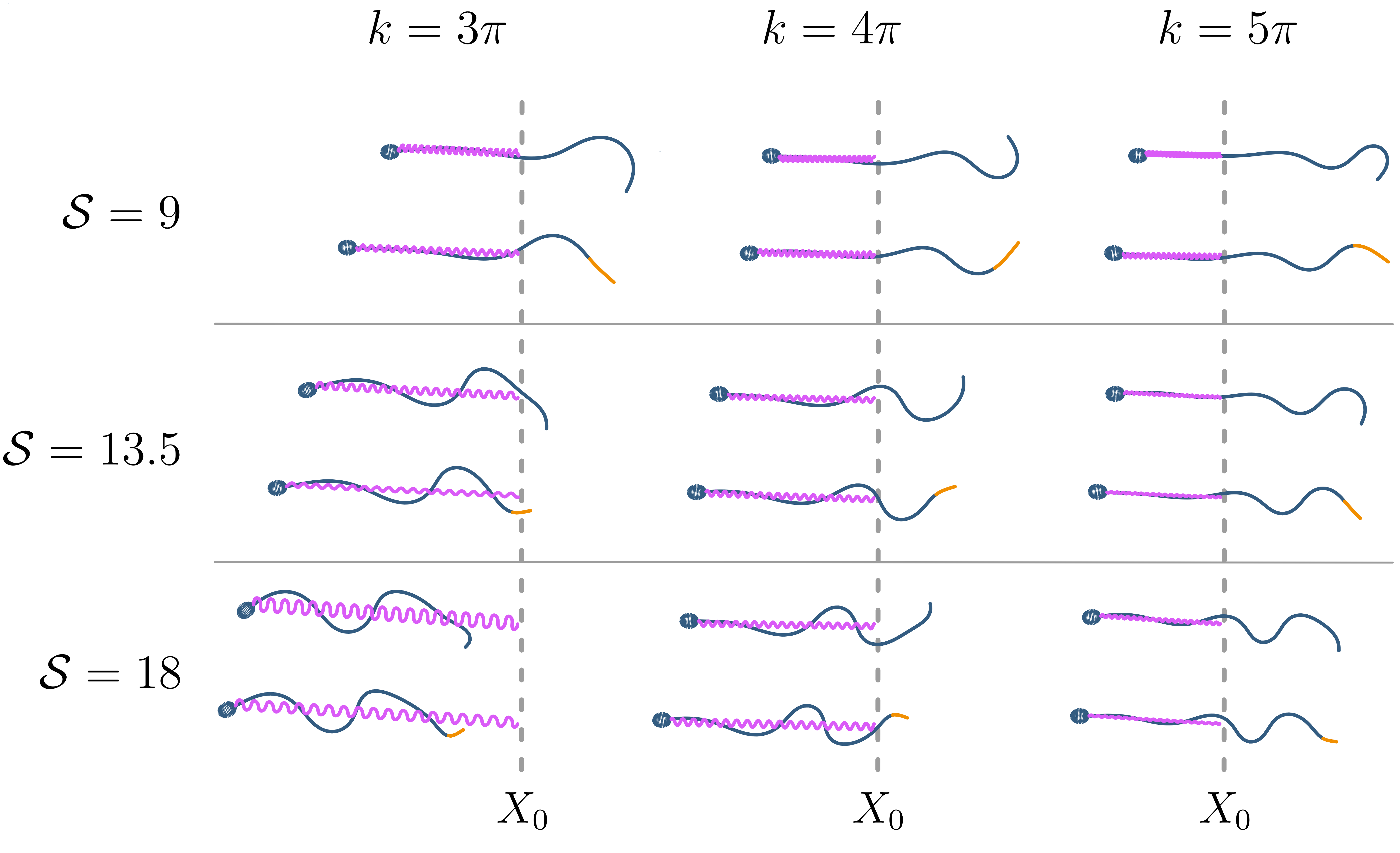}
    \caption{
         The swimming track following the head for both fully-active and optimally-inactive sperm (top and bottom of each pair respectively) for a selection of parameter pairs (\(\mathcal{S},k\)) actuated with \(\mathcal{M}=\mathcal{M}_\eta\) and simulated for 10 cycles of the active moment. For each pair the sperm are plotted at the final time point with the optimally-inactive length (\(\ell=\ell_{\eta}\)) shown in orange.
    }
    \label{fig:tracks_Meta}
    \end{figure}
    
\subsection{Simulations actuated with $\mathcal{M} =\mathcal{M}_\eta$}
\label{sec:results_meta}
In addition to the figures in Sec.~\ref{sec:results_mval}, equivalent results can be produced for cells actuated with $\mathcal{M}=\mathcal{M}_\eta$. 
Figures~\ref{fig:val_eta_Meta}a,\,b show the relationship between active tail length \(\ell\) and swimming speed and efficiency (with finer detail shown in Fig.~\ref{fig:optimallength_Meta}a,\,b). Figures~\ref{fig:val_eta_Meta}c\,d highlight the relative benefit of an optimally-inactive flagellum compared to the fully-active case, showing an increase of 7--40\% for swimming speed and 22--240\% for efficiency. The optimum active length for both swimming speed and efficiency is shown in Figs.~\ref{fig:optimallength_Meta}c,\,d for $k\in[3\pi,5\pi]$ and $\mathcal{S}\in[9,18]$. 
The qualitative similarities between Figs.~\ref{fig:val_eta_Meta} and \ref{fig:optimallength_Meta}, and the analogous figures in Sec.~\ref{sec:results_mval} (Figs.~\ref{fig:results-main} and \ref{fig:optimallength}) highlight that the observed cell behaviors are not simply governed by the choice of actuation parameter \(\mathcal{M}\).

\begin{figure}[t]
    \centering
    \includegraphics[width=\textwidth]{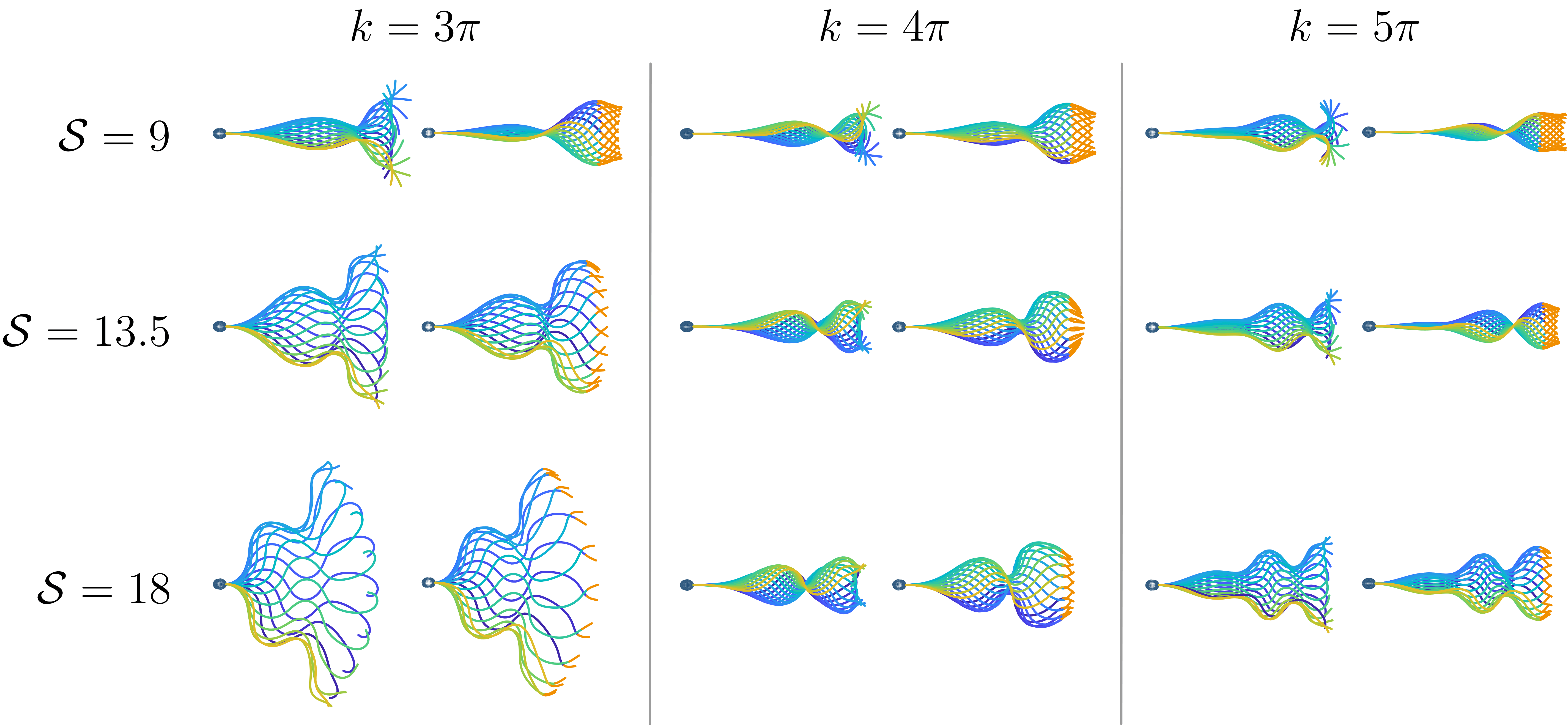}
    \caption{
         The flagellar waveform for a fully-active and optimally-inactive sperm (left and right of each pair respectively) simulated for a selection of parameter pairs (\(\mathcal{S},k\)), and actuated with \(\mathcal{M}=\mathcal{M}_\eta\). The color of the flagellum indicates progression through a single beat, from blue to yellow. For each pair the sperm are plotted with the optimally-inactive length \(\ell=\ell_{\eta}\), shown in orange.
    }
    \label{fig:waveforms_Meta}
\end{figure}

\subsection{Effect of the end piece on waveform and trajectory}

The foregoing results establish the propulsive and efficiency advantages of an inactive distal flagellar region. 
To better understand why the inactive region yields these advantages we now consider the flagellar waveforms and overall cell dynamics resulting from simulations with parameters \(\mathcal{S}=\{9, 13.5, 18\}\) and \(k=\{3\pi, 4\pi, 5\pi\}\). The tracks and shapes of swimming sperm actuated with \(\mathcal{M}=\mathcal{M}_\text{VAL}\) are shown in Fig.~\ref{fig:tracks}, which compares the fully-active (\(\ell=1\)) and optimally-inactive ($\ell=\ell_\text{VAL}$) waveforms for each parameter pair. Simulations with an optimally-inactive region produce flagellar shapes and tracks that are more qualitatively `sperm-like' than those with an entirely active flagellum, and more specifically exhibit lower curvature and hence tangent angle in the distal flagellum. The change to the flagellar envelope is most clearly visible in the overlaid timelapse images of Fig.~\ref{fig:waveforms}. A similar effect is observed for cells actuated with $\mathcal{M}=\mathcal{M}_\eta$, whose cell tracks can be seen in Fig.~\ref{fig:tracks_Meta}, and waveforms in Fig.~\ref{fig:waveforms_Meta}.

\begin{figure}[t]
    \centering
    \includegraphics[width=\textwidth]{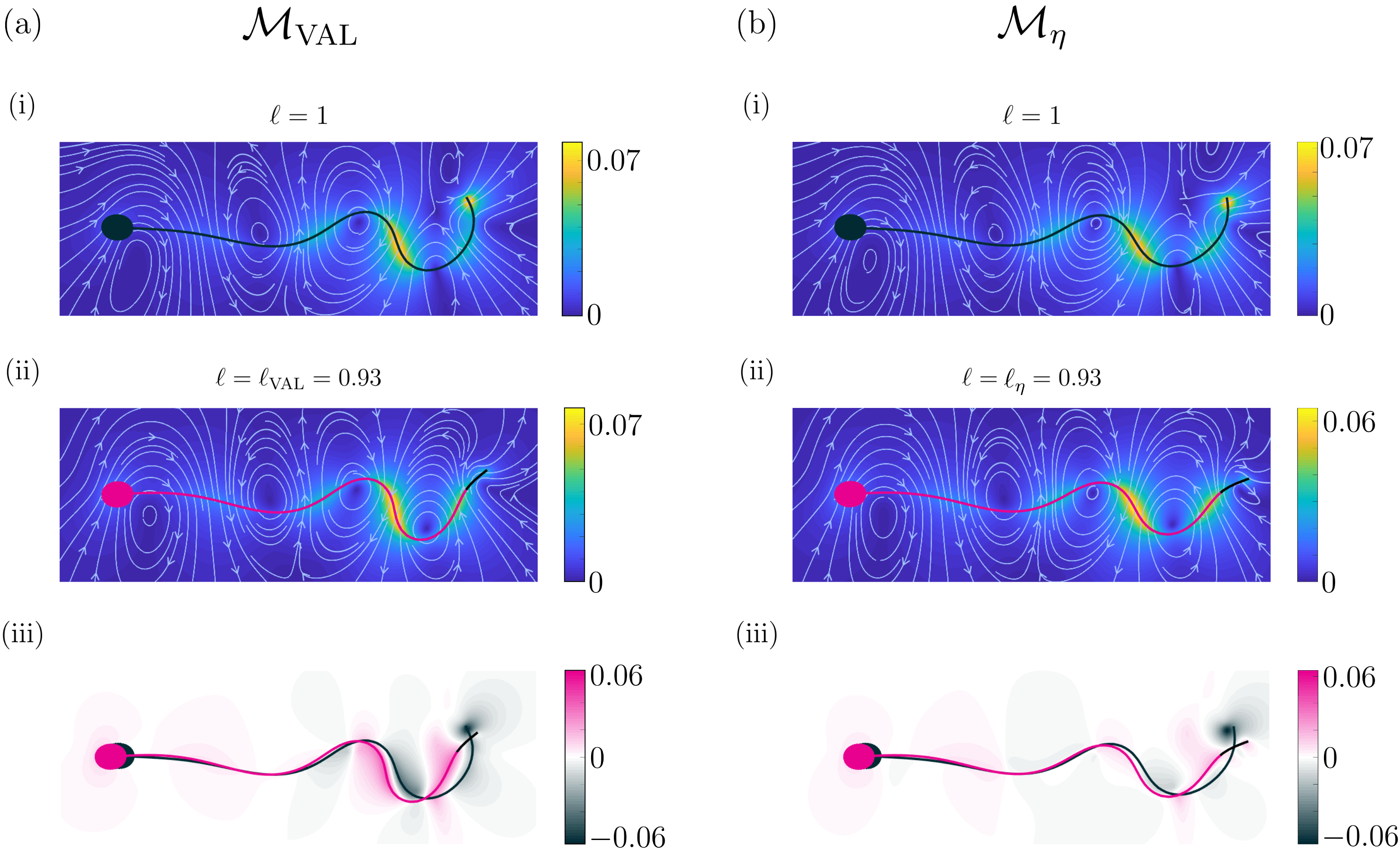}
    \caption{
        Comparison of the flow fields around a simulated sperm at \(t=8\pi\) with \(\mathcal{S} = 13.5\) and \(k = 4\pi\) for (a) a cell actuated with \(\mathcal{M}=\mathcal{M}_\text{VAL}\) and (b) \(\mathcal{M}=\mathcal{M}_\eta\). In each case we plot (i) a cell with a fully-active flagellum, (ii) an optimally-inactive cell and (iii) the difference between flow fields, with positive values corresponding to the optimally-inactive case having a faster fluid velocity. (i,\,ii) Instantaneous fluid streamlines are shown in white with arrows indicating direction, and the fluid magnitude is indicated by the colorbar for each panel. In (ii,\,iii), the inactive part of the flagellum is shown in black. For additional time points see Appendix~\ref{app:flow}. 
    }
    \label{fig:results-streamlines}
\end{figure}

The velocity field associated with flagella that are fully-active (\(\ell=1\)) and optimally-inactive (\(\ell=\ell_{\text{VAL}}\)) for propulsion are shown in Fig.~\ref{fig:results-streamlines}a at $t=8\pi$ for parameter values $\mathcal{M}=\mathcal{M}_\text{VAL}$, $\mathcal{S}=13.5$ and $k=4\pi$. The qualitative features of both waveform and the velocity field are similar, however the optimally-inactive flagellar waveform has reduced curvature and tangent angle in the distal region, resulting in an additional `oblique' region (i.e.\ where \(\theta-\phi \approx \pi/4\)) that confers additional thrust to the cell when \(\ell < 1\). The equivalent results for cells actuated with \(\mathcal{M}=\mathcal{M}_\eta\) can be seen in Fig.~\ref{fig:results-streamlines}b. Plots at additional time points are given in Appendix~\ref{app:flow} for both values of \(\mathcal{M}\).

\section{Discussion} 
In simulations, we observe that spermatozoa which feature a short, inactive region at the end of their flagellum swim faster and more efficiently than those without. 
For each simulation, cell motility is optimized when 2--18\% of the distal flagellum is inactive, regardless of parameter choices.
For the larger choices of \(\mathcal{S}\), commonly seen in the human female reproductive tract, the optimally inactive length for velocity and efficiency (Figs.~4c and 6d) lies between 2--10\%.
Experimental measurements of human sperm indicate an average combined length of the midpiece and principal piece of \( \approx 54\,\mu \)m and an average end piece length of \( \approx 3\,\mu \)m \cite{cummins1985mammalian}, suggesting that the effects uncovered in this work are biologically important. 

The modeling method has been validated by comparing the mean absolute value of curvature along the flagellum, for the case \(\mathcal{S} = 18,\ k=5\pi\), with that of an experimentally captured sperm (details provided in Appendix~\ref{app:stiff}). 
The consistency between simulated cells and experimental data lends further confidence that the modeling choices made are biophysically reasonable; future investigation may assess the robustness of the conclusions to more intricate models of flagellar structure and regulation.


Sperm move through a variety of fluids during migration, in particular encountering a step change in viscosity when penetrating the interface between semen and cervical mucus, and having to swim against physiological flows \cite{Tung2015}.
Cells featuring an optimally-sized inactive end piece may form better candidates for fertilization, being able to swim faster and for longer when traversing the female reproductive tract \cite{holt2015sperm}.

The basic mechanism by which the flagellar wave produces propulsion is through the interaction of segments of the filament moving obliquely through the fluid \cite{gray1955propulsion}, with the angle relative to the direction of propulsion being between \(-\pi/2\) and \(\pi/2\). Analysis of the flow fields (Fig.~\ref{fig:results-streamlines}; Appendix~\ref{app:flow}) suggest that the lower curvature and hence tangent angle associated with the inactive end piece maintains such regions towards the end of the flagellum, preventing the flagellum from `overturning' (Figs.~\ref{fig:tracks} and \ref{fig:tracks_Meta}). 

An inactive region of flagellum is not a feature unique to human gametes -- its presence can also be observed in the sperm of other species \cite{fawcett1975mammalian}, as well as other microorganisms.  
In particular, the axonemal structures of the biflagellated algae \textit{Chlamydomonas reinhardtii} deplete at the distal tips \cite{jeanneret2016}, suggesting the presence of an inactive region.
The contribution to swimming speed and cell efficiency due to the inactive distal section in these cases remains unknown.
By contrast, the tip of the 9+2 cilium is a more organized ``crown" structure \cite{kuhn1978structure}, which will interact differently with fluid than the flagellar end piece modeled here. The structure and actuation of the motile 9+0 cilia found in the embryonic node (see for example \cite{nonaka1998}) is less well-characterized.
Understanding this distinction between cilia and flagella, as well as the role of the inactive region in other microorganisms, may provide further insight into underlying biological phenomena, such as chemotaxis \cite{ALVAREZ2014198} and synchronization \cite{guo2018bistability, goldstein2016elastohydrodynamic}. Further work should investigate how this phenomenon changes when more detailed models of the flagellar ultrastructure are considered, taking into account the full 9+2 structure \cite{ishijima2019modulatory}, sliding resistance associated with filament connections \cite{coy2017counterbend}, and the interplay of these factors with biochemical signaling in the cell \cite{carichino2018}.

The ability to qualitatively assess and model the inactive end piece of a human spermatozoon could have important clinical applications. In live imaging for diagnostic purposes, the end piece is often hard to resolve due to its depleted axonemal structure. Lacking more sophisticated imaging techniques, which are often expensive or impractical in a clinical environment, modeling of the end piece combined with flagellar tracking software, such as FAST \cite{gallagher2019rapid}, could enable more accurate sperm analysis, and help improve cell selection in assisted reproductive technologies. The difficulty in capturing the end piece may be a significant explanatory factor to why previous works have only been able to reconstruct the qualitative features of the swimming tracks, even when high quality flagellar waveform data is available \cite{ishimoto2017coarse}. Furthermore, knowledge of the function of an inactive distal region has wider applications across synthetic microbiology, particularly in the design of artificial swimmers \cite{dreyfus2005microscopic} and flexible filament microbots used in targeted drug delivery \cite{montenegro2018microtransformers}. 

\section{Summary and Conclusions} 
In this paper, we have revealed the propulsive and energetic advantages conferred by an inactive distal region of a unipolar ``pusher'' actuated elastic flagellum, characteristic of mammalian sperm. The optimal inactive flagellum length depends on the balance between elastic stiffness and viscous resistance, and the wavenumber of actuation. The optimal inactive fraction mirrors that seen in human sperm ($\approx 3\,\mu$m$/57\,\mu$m, or \( \approx 5\% \)). From a modeling point of view, inclusion of an inactive region can radically change the waveform, propulsive velocity and efficiency, and so may be crucial to include in biophysical studies. These findings also motivate the development of more highly-resolved methods to image the far distal flagellum.
Furthermore, inclusion of an inactive region may be an interesting avenue to explore when improving the efficiency of artificial microswimmer designs. Finally, important biological questions may now be posed; for example does the presence of the inactive end piece confer an advantage to cells penetrating highly viscous cervical mucus?

\section*{Acknowledgments} 
D.J.S. and M.T.G. acknowledge funding from the Engineering and Physical Sciences Research Council (EPSRC) Healthcare Technologies Award (EP/N021096/1). M.T.G. acknowledges support of the University of Birmingham through its Dynamic Investment Fund. C.V.N. and A.L.H-M. acknowledge support from the EPSRC for funding via PhD scholarships (EP/N509590/1). J.C.K-B. acknowledges the National Institute of Health Research (NIHR) and Health Education England, Senior Clinical Leadership Grant: The role of the human sperm in healthy live birth (NIHRDH-HCS SCL-2014-05-001). This article presents independent research funded in part by the NIHR and Health Education England. The views expressed are those of the authors and not necessarily those of the NHS, the NIHR or the Department of Health. The authors also thank Hermes Gad\^elha (University of Bristol, UK), Thomas D. Montenegro-Johnson (University of Birmingham, UK) for stimulating discussions around elastohydrodynamics and Gemma Cupples (University of Birmingham, UK) for the experimental still in Fig.~\ref{fig:experimental} of Appendix~\ref{app:stiff} (provided by a donor recruited at Birmingham Women's and Children's NHS Foundation Trust after giving informed consent). A significant portion of the computational work described in this paper was performed using the University of Birmingham's BlueBEAR HPC service, which provides a High Performance Computing service to the University's research community. See \url{http://www.birmingham.ac.uk/bear} for more details.

\newpage
\appendix

\newpage
\section{Qualitative validation using experimental data}
\label{app:stiff}
\begin{figure}[ht]
    \centering
    \includegraphics[width=0.8\textwidth]{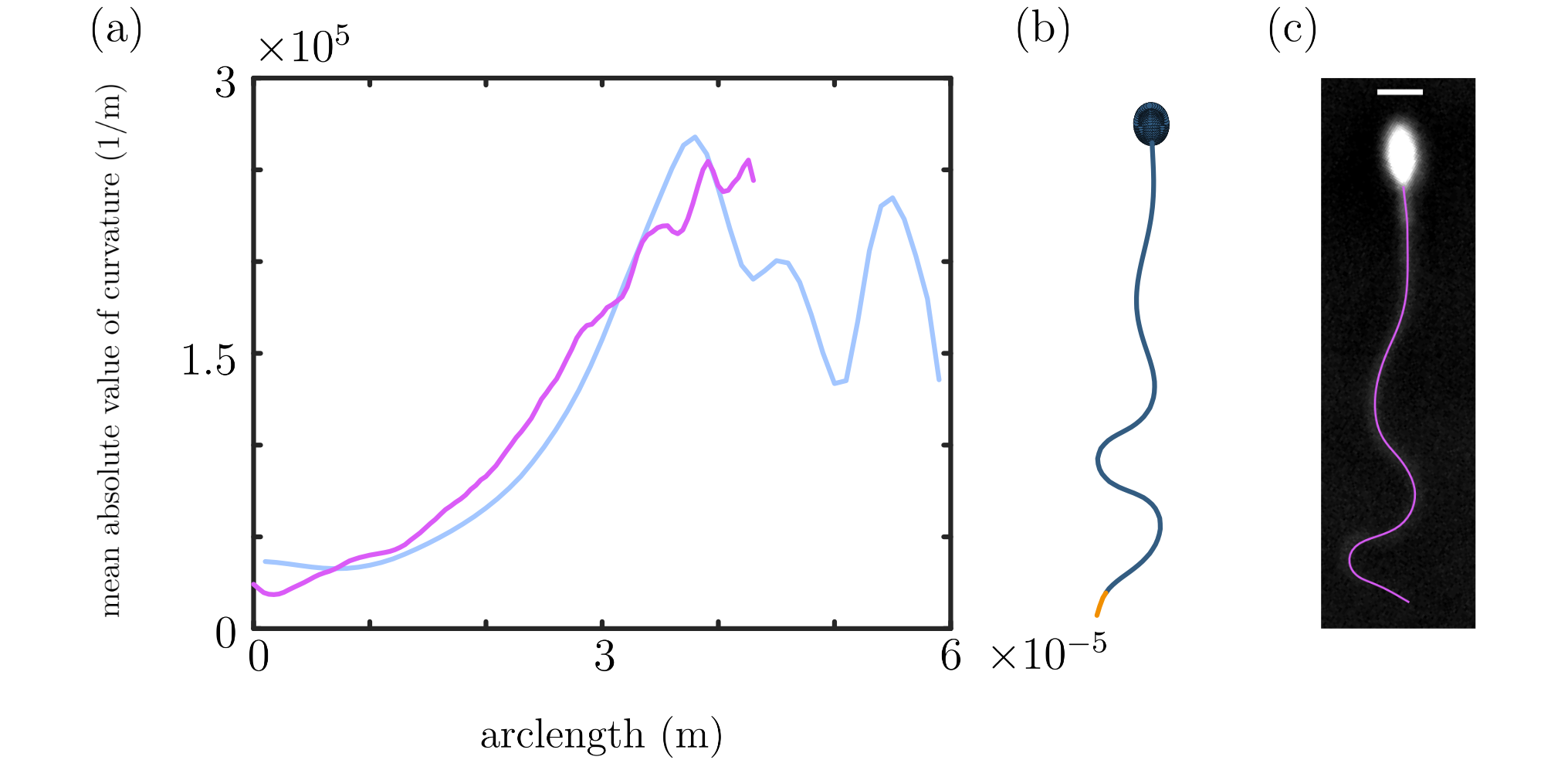}
    \caption{
        Comparison between simulated and experimental data to verify the choice of elastic stiffness parameters. (a) The mean absolute value of curvature for both a cell simulated with {\((\mathcal{S},k)~=~(18,5\pi)\)} (blue) and an experimentally tracked sperm (magenta) is plotted against arclength. A waveform comparison is shown for the (b) simulated and (c) experimental cell (tracked with FAST  \cite{gallagher2019rapid}, scale bar denotes $5\,\mu$m). In the case of the simulated waveform, the inactive end piece is shown in orange.
    }
    \label{fig:experimental}
\end{figure}

The proximal-distal stiffness ratio \(\rho = 36.4\)  was chosen to provide results that closely resemble those waveforms seen in swimming human spermatozoa. To validate this choice, and also assess the internal active moment and stiffness model~(Eq.~\eqref{eq:stiffness}), the mean absolute value of curvature is plotted against arclength of both a simulated (with \((\mathcal{S},k) = (18,5\pi)\)) and experimental cell in Fig.~\ref{fig:experimental}.  The experimental data and waveform agree closely over the first \(40\,\mu\)m of flagellum, while the remainder of the flagellum is not visible in the experimental image. To emphasize, this study is not focusing on parameter estimation or improving fitting of flagellar waveforms, rather the aim is to find indicative parameters that are capable of broadly matching experimental data and hence enable a rational exploration of the inactive end piece effect.

\newpage
\section{Additional flow fields}
\label{app:flow}
\begin{figure}[h]
    \centering
    \includegraphics[width=\textwidth]{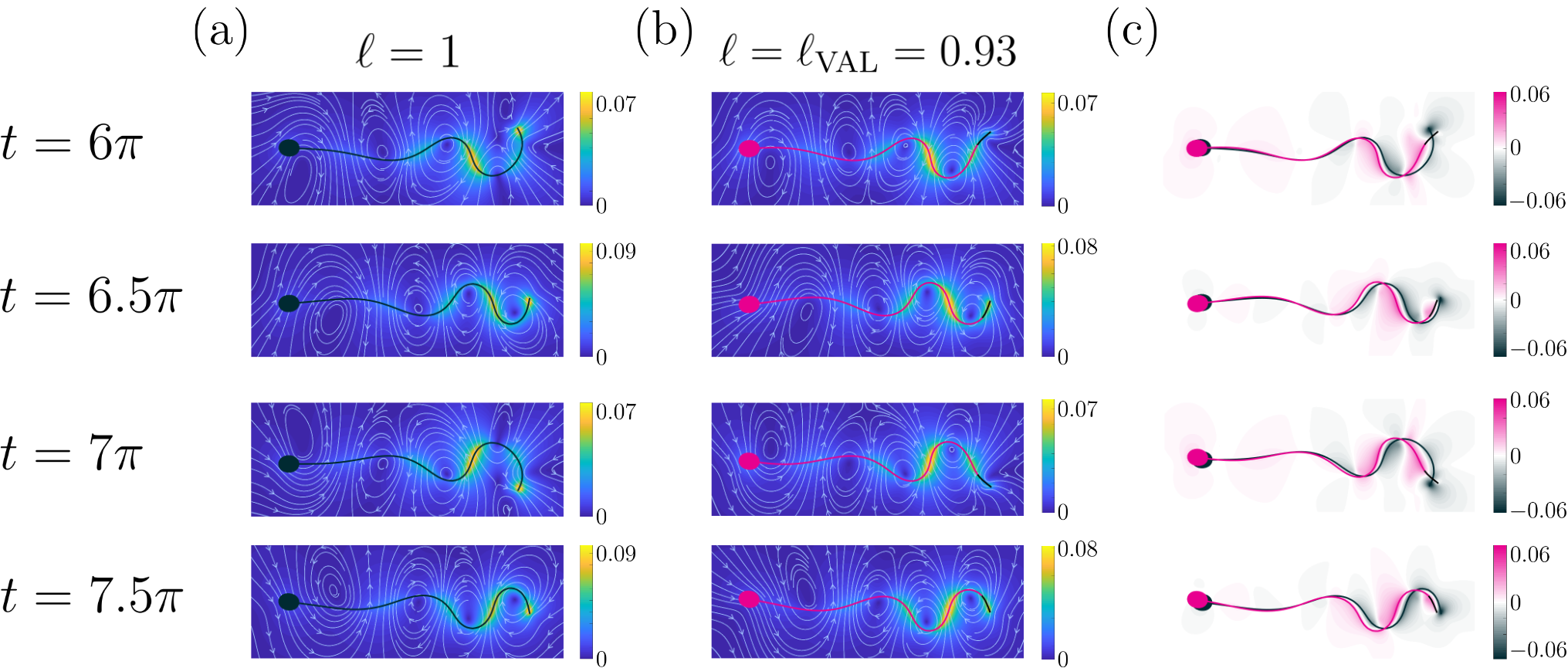}
    \caption{
        Comparison of the flow fields around a simulated sperm actuated with \(\mathcal{M}=\mathcal{M}_\text{VAL}\) at \(t=8\pi\) with \(\mathcal{S} = 13.5\) and \(k = 4\pi\), for (a) a cell with a fully-active flagellum, and (b) an optimally-inactive cell, each shown at four time points in the final simulated beat. (c) The difference between flow fields, with positive values corresponding to the optimally-inactive case having a faster fluid velocity. (a,\,b) Instantaneous fluid streamlines are shown in white with arrows indicating direction, and the fluid magnitude in indicated by the colorbar for each panel. In (b) the inactive part of the flagellum is shown in black.
    }
    \label{fig:additional_flow}
\end{figure}

\begin{figure}[h]
    \centering
    \includegraphics[width=\textwidth]{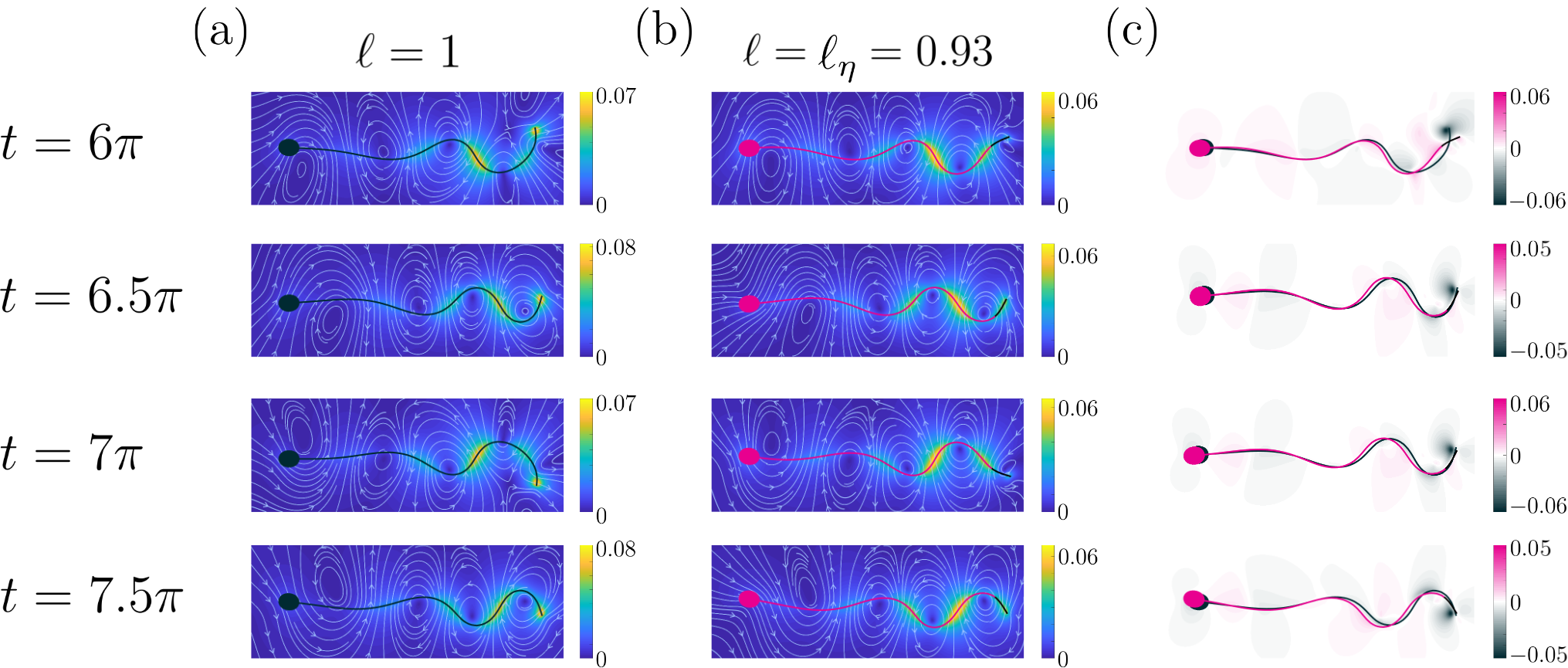}
    \caption{
        Comparison of the flow fields around a simulated sperm actuated with \(\mathcal{M}=\mathcal{M}_\eta\) at \(t=8\pi\) with \(\mathcal{S} = 13.5\) and \(k = 4\pi\), for (a) a cell with a fully-active flagellum, and (b) an optimally-inactive cell, each shown at four time points in the final simulated beat. (c) The difference between flow fields, with positive values corresponding to the optimally-inactive case having a faster fluid velocity. (a,\,b) Instantaneous fluid streamlines are shown in white with arrows indicating direction, and the fluid magnitude in indicated by the colorbar for each panel. In (b) the inactive part of the flagellum is shown in black.
    }
    \label{fig:additional_flow_Meta}
\end{figure}

Additional flow fields for simulated sperm with $\mathcal{S}=13.5$, $k=4\pi$ and actuated with $\mathcal{M}=\mathcal{M}_\text{VAL}$ are shown in the case of a fully active flagellum and an optimally-inactive flagellum in Fig.~\ref{fig:additional_flow}. The analogous plots for $\mathcal{M}=\mathcal{M}_\eta$ can be seen in Fig.~\ref{fig:additional_flow_Meta}.

\newpage
\bibliographystyle{unsrt}

\end{document}